\documentclass[a4paper,11pt,twocolumn]{article} 

\usepackage{graphicx} 
\usepackage{bbold} 
\usepackage{mathtools}
\usepackage{amsmath}
\numberwithin{equation}{section} 
\usepackage{amsfonts}
\usepackage{amssymb} 
\usepackage{slashed} 
\usepackage{tabularx}
\usepackage{cite}
\usepackage{color} 
\usepackage[dvipsnames]{xcolor}
\definecolor{KBFIred}{RGB}{163,35,47}
\definecolor{linkblue}{RGB}{7, 140, 255}
\usepackage{url}
\usepackage{hyperref}
\hypersetup{
	colorlinks = true,
	allcolors = linkblue} 
\usepackage{bm}
\usepackage{mathrsfs} 
\usepackage{framed}

\usepackage[font=footnotesize,labelfont=bf]{caption}
\def\eq#1{{Eq.~(\ref{#1})}}

\newcommand{\be}{\begin{equation}}
\newcommand{\ee}{\end{equation}}
\newcommand{\bea}{\begin{eqnarray}}
\newcommand{\eea}{\end{eqnarray}}

\newcommand*\xbar[1]{%
  \hbox{\;%
    \vbox{%
      \hrule height 0.5pt 
      \kern0.5ex
      \hbox{%
        \kern-0.25em
        \ensuremath{#1}%
        \kern-0.07em
      }%
    }%
  }%
} 
\newcommand{\com}[1]{}
\newcommand{\gsim}{\lower.7ex\hbox{$\;\stackrel{\textstyle>}{\sim}\;$}}
\newcommand{\lsim}{\lower.7ex\hbox{$\;\stackrel{\textstyle<}{\sim}\;$}} 

\newcommand{\bc}{\begin{center}}
\newcommand{\ec}{\end{center}}

\newcommand{\ES}{\Lambda}

\newcommand{\alphaD}{\alpha_{\scriptscriptstyle{D}}}
\newcommand{\eD}{e_{\scriptscriptstyle{D}}}
\newcommand{\AD}{A_{\scriptscriptstyle{D}}}
\newcommand{\FD}{F_{\scriptscriptstyle{D}}}
\newcommand{\gammaD}{\gamma_{\scriptscriptstyle{D}}}
\newcommand{\epsilonD}{\varepsilon_{\scriptscriptstyle{D}}}
\newcommand{\alphaEW}{\alpha_{\scriptscriptstyle{EW}}}

\newcommand{\pB}{p_{\scriptscriptstyle B}}
\newcommand{\pK}{p_{\scriptscriptstyle K}}
\newcommand{\SM}{\scriptscriptstyle {\rm SM}}
\newcommand{\SF}{\scriptscriptstyle {\rm SF}}

\newcommand{\mKs}{m_{K^{*}}}
\newcommand{\mB}{m_{B}}

\font\beeg=cmr17 scaled 1800
\newbox\ibox
\def\versal#1{\setbox\ibox=\hbox{{\beeg #1}~}%
	    \noindent\global\hangindent=\wd\ibox\global\hangafter-2%
	    \sc\smash{\llap {\lower 14pt \box\ibox}}}

\begin{document}
\onecolumn
\thispagestyle{empty}
\begin{center}
{ \Large \color{KBFIred} \textbf{ 
    Explaining the $B^+\to K^+ \nu \bar{\nu}$ excess via a
    massless dark photon \\[0.3cm]
}}
\vspace*{1.5cm}
{
{\bf E. Gabrielli$^{{a,b,c}}$, L. Marzola$^{{c}}$, K. M\"u\"ursepp$^{{c}}$ and }
 {\bf M. Raidal$^{{c}}$}
}\\

\vspace{0.5cm}
{\small 
  {\it 
    (a) Physics Department, University of Trieste, Strada Costiera 11, \\ I-34151 Trieste, Italy}
    \\[1mm]
{\it   (b)
INFN, Sezione di Trieste, Via Valerio 2, I-34127 Trieste, Italy}
\\[1mm]
    {\it 
(c) Laboratory of High-Energy and Computational Physics, NICPB, R\"avala Pst. 10, \\ 10143 Tallinn, Estonia}
}
\ec

\vskip3cm
\bc
{\bf ABSTRACT}
\ec

\vspace*{5mm}

\noindent 
The Belle II collaboration has recently observed the rare decay $B^+\to K^+ \nu \bar{\nu}$, finding an excess with respect to the Standard Model prediction. We explore the possibility that the data entails long-distance interactions induced by a massless dark photon, $\gammaD$. This couples at the tree-level to an invisible, dark sector and to the Standard Model via higher-dimensional operators, such as the chromomagnetic-dipole coupling that we use to explain  the excess. As the process $B^+\to K^+ \gammaD$ is forbidden by angular momentum conservation, the transition mediated by the off-shell dark photon yields a three-body final state comprising a pair of dark fermions that show as a missing energy continuum in the detector, faking the neutrino signature. We show that the Belle II data is explained for perturbative values of the parameters of the model. This scenario predicts new contributions to the neutral $B$ meson decays $B^0\to K^* \gammaD$, in which the emission of a on-shell dark photon is allowed, yielding a monochromatic missing energy signature. Analogously, an excess due to the emission of a dark photon is predicted for the $B^0_s\to \phi + E_{\rm miss}$ decay that could be scrutinized next at the LHCb experiments.

\newpage

\section{Introduction\label{sec:intro}}
The flavor changing neutral current (FCNC) processes showing in the $B$ meson decays are a powerful tool for new physics (NP) searches. In particular, the semi-leptonic decays proceeding at the quark level through the $b\to s \nu \bar{\nu}$ transition offer an excellent arena to test the Standard Model (SM) predictions. Due to the absence of FCNC at the tree-level in the SM, these processes are also inherently sensitive to any NP interacting with quarks and leptons and, consequently, quite useful to constrain possible radiative effects sourced by heavy states beyond the SM.

Recently, the Belle II collaboration has measured the branching ratio (BR) related to the rare decay $B^+\to K^+ \nu \bar{\nu} $ observed with a $3.5\sigma$ significance. The observation relied on two different strategies for measuring the BR: the standard hadronic-tag (had) method and the novel inclusive-tag (incl) technique. The branching ratios obtained are~\cite{Belle-II:talk1,Belle-II:2023esi} 
\bea
    {\rm BR}(B^+\to K^+ \nu \bar{\nu})_{\rm had}&=&
    \left(1.1_{-0.8-0.5}^{+0.9+0.8}\right)\times 10^{-5}\, ,
      \\
    {\rm BR}(B^+\to K^+ \nu \bar{\nu})_{\rm incl}&=&
    \left(2.7\pm 0.5 \pm 0.5\right)\times 10^{-5}\, ,
\eea
where the quoted uncertainties correspond to the statistical and systematic errors, respectively. The combination of these measurements gives
\be
    {\rm BR}(B^+\to K^+ \nu \bar{\nu})_{\rm exp}=
    \left(2.3\pm 0.7 \right)\times 10^{-5}\, ,
\label{BKexp}
    \ee
to be compared with the corresponding SM prediction~\cite{Altmannshofer:2009ma,Buras:2014fpa}
\be
    {\rm BR}(B^+\to K^+ \nu \bar{\nu})_{\rm SM}=
    \left(4.29\pm 0.23\right)\times 10^{-6}\, ,
\label{BKSM}
    \ee
where the tree-level contribution from $B^+\to \tau^+(\to K^+\bar{\nu})\nu$ was subtracted.

As we can see, the BR measurement~\eqref{BKexp} shows a $2.7\sigma$ discrepancy when compared with the SM prediction~\eqref{BKSM}. The mismatch is driven by the inclusive-tag result and, since the hadronic-tag value is consistent with the SM expectation within errors, it could well be due to unknown uncertainties pertaining to the new method; future measurements will certainly clarify this issue. 

In the meantime, one may wonder if the signal shows a preliminary hint of physics beyond the SM. In this case, the corresponding contribution to the BR indicated by the experiment is
\be
    {\rm BR}(B^+\to K^+ \nu \bar{\nu})_{\rm NP}=
    \left(1.9\pm 0.7\right)\times 10^{-5}\, .
\label{BKNP}
\ee

The possible deviations of semi-leptonic $B$ decays with neutrino final states from the SM predictions stem from two distinct classes of NP contributions.
The first one, associated to the so called {\it indirect}-NP effects, is related to the presence of heavy NP particles that add to the SM Wilson coefficients of the interaction Hamiltonian describing the $B$ semi-leptonic decays in the SM effective field theory approach. The other one is related to the so called {\it direct}-NP effects, and is linked to the presence of new invisible particles(antiparticles) $X(\bar{X})$ that could be produced via two-body $B^+\to K^+ X$  or three-body $B^+\to K^+ X \bar{X}$ decays, depending on the spin. When these particle are stable (or long-lived enough to escape the detector), they mimic the neutrino missing energy signature expected for $B^+\to K^+ \nu \bar{\nu}$ within the SM. Several recent studies have suggested NP interpretations of the Belle II result, both in the framework of indirect~\cite{Athron:2023hmz, Bause:2023mfe,Allwicher:2023xba,He:2023bnk,Datta:2023iln,Gorbahn:2023juq,Chen:2024jlj} and direct NP effects~\cite{Gorbahn:2023juq,Bruggisser:2023npd, Felkl:2023ayn,Berezhnoy:2023rxx,Altmannshofer:2023hkn,Fridell:2023ssf,McKeen:2023uzo}. 

In order to explain the Belle II result, we focus here on the latter possibility ascribing the excess to a pair of light dark fermions, $Q_i$, produced in the $B^+\to K^+ Q_i\bar{Q}_i$  transition via long-distance interactions mediated by a massless dark photon $\gammaD$. On general grounds, it is also expected that the same new FCNC interactions give contributions to the $B^0\to K^*$ transition, which presently is consistent with the SM predictions for $B^0\to K^* \nu\bar{\nu}$.  

Massless dark photons are the quanta associated with an unbroken $U(1)_D$ gauge symmetry of a hypothetical dark sector, comprising particles that do not partake in the SM interactions. The dark photon scenario~\cite{Fayet:1980rr,Okun:1982xi,Georgi:1983sy,Holdom:1985ag,Fayet:1990wx} has been extensively analyzed in the Literature, mainly in its massive limit, and is the subject of many current experimental searches~\cite{Alexander:2016aln} -- see~\cite{Fabbrichesi:2020wbt} for a recent review. Massless dark photons have been explored in the context of astrophysics and cosmology as sources of long range interactions (``dark electromagnetism'') among the dark matter constituents~\cite{Gradwohl:1992ue,Carlson:1992fn,Foot:2004pa,Ackerman:2008kmp,Fan:2013tia,Foot:2014uba,Heikinheimo:2015kra,Agrawal:2016quu,Acuna:2020ccz}.
Massless dark photons have also been employed to explain the SM flavor hierarchy puzzle~\cite{Gabrielli:2013jka,Gabrielli:2016vbb} and the  experimental tests of this scenario involving Higgs physics~\cite{Gabrielli:2014oya,Biswas:2015sha,Biswas:2016jsh,Biswas:2017lyg,Biswas:2022tcw,Beauchesne:2023bcy,ATLAS:2022xlo,CMS:2020krr}, flavor-changing neutral currents~\cite{Gabrielli:2016cut}, and kaon physics~\cite{Fabbrichesi:2017vma}.

Unlike the massive case, a massless dark photon does not have tree-level interactions with ordinary matter and, for this reason, is much less constrained by experiments~\cite{Holdom:1985ag,Fabbrichesi:2020wbt}. Contact with the SM is provided by effective couplings sourced at the loop level by heavy messenger fields, which connect the SM and the dark sector. For the SM fermions, the lowest-order coupling with the dark photon is provided by the magnetic-dipole operator~\cite{Fabbrichesi:2020wbt,Dobrescu:2004wz} that could also induce FCNC transitions~\cite{Gabrielli:2016cut}. Hence, assuming that the dark sector comprises $N_Q$ light dark fermions enjoying an unbroken $U(1)_D$ symmetry (a dark sector replica of QED), we study the inclusive decay $B^+\to K^+ Q_i\bar{Q_i}$ due to the ``dark'' chromomagnetic-dipole operator for the $b\to s \gammaD$ ~\cite{Gabrielli:2016cut} process, including the relevant Sommerfeld-Fermi corrections~\cite{Isidori:2007zt}. The decay $B^+\to K^+ \gammaD$ is precluded by angular momentum conservation for a real massless dark photon. Consequently, the long-distance process underlying the transition must necessarily involve a virtual state yielding a dark fermion pair in the final state and the resulting $B^+\to K^+ Q_i\bar{Q_i}$ decay can be used to explain the Belle II excess ascribed to $B^+ \to K^+ \nu \bar{\nu}$. Differently, the emission of a real dark photon is allowed in the related $B^0\to K^* \gammaD$ decay, which can be used to constrain the scenario since no events have been observed so far for this process. For the same reason, the scenario also entails the presence of a NP signal showing as $B^0_s\to \phi + E_{\rm miss}$ at dedicated searches and potentially observable already at the LHCb experiment.  

The paper is organized as follows. In section~\ref{sec:th} we provide the theoretical framework describing the dark photon scenario and the relevant FCNC couplings. In sections~\ref{sec:BK} and~\ref{sec:BKstar} we provide analytical and numerical results for the branching ratios of $B^+\to K^+ Q_i\bar{Q}_i$ and $B^0\to K^* \gammaD$, respectively. In section~\ref{sec:BKanom} we analyze the Belle II excess
and the constraints due to the $B^0\to K^* \nu \bar{\nu}$ process. In section~\ref{sec:fpredict}, instead, we provide the predictions of the scenario for the
$B^0_s\to \phi \nu \bar{\nu}$ decay. Finally, our conclusions are given in section \ref{sec:concl}.
  
\section{Theoretical framework\label{sec:th}}

In this section we analyze the Belle II excess within the framework of a NP scenario that leverages the phenomenology of dark photons.

Broadly speaking, dark photon models can be categorized into two distinct classes depending on whether the corresponding field quanta are massive or massless. The choice yields very different experimental signatures: a massless dark photon does not possess tree-level couplings to any SM current and is thus allowed to interact with ordinary matter only through operators of dimension higher than four \cite{Holdom:1985ag,Dobrescu:2004wz,Fabbrichesi:2020wbt}. Differently, massive dark photons can couple to ordinary matter via a renormalizable operator of dimension four involving an arbitrarily small charge \cite{Holdom:1985ag}. The two classes are disjoint, in that the massless case cannot be obtained as a limiting case of massive dark photon models. For the sake of explaining the Belle II excess, we will resort to a massless dark photon.

In order to describe the FCNC interactions causing the $b\to s$ transition of interest, we use the following effective Lagrangian involving $b$ and $s$ quarks 
\be
   {\cal L}_{eff}=\frac{1}{2\ES} \left[\bar{s}\sigma_{\mu\nu} b\right] \FD^{\mu\nu} + h.c.\, ,
   \label{Leff}
\ee
where $\sigma_{\mu\nu}=i/2[\gamma_{\mu},\gamma_{\nu}]$, and the $\FD^{\mu\nu}=\partial^{\mu} \AD^{\nu} - \partial^{\nu} \AD^{\mu}$ is the dark photon field strength tensor. The effective scale $\ES$ depends on the couplings and masses of the dark and messenger sectors as specified in Ref.~\cite{Gabrielli:2016cut}.
In addition, we assume that the dark photon is coupled to dark fermions in a way reminiscent of QED within the SM,
\be
   {\cal L}_{dark}=\eD \sum_i q_i \left[\bar{Q}_i\gamma_{\mu}Q_i \right] \AD^{\mu}\, ,
\label{Ldark}
\ee
where the index $i$ indicates the dark fermion generation, $\eD$ is the coupling of the $U(1)_D$ group with the corresponding dark fine-structure constant $\alphaD=\eD^2/4\pi$, and $q_i$ are the corresponding charges of the dark fermions $Q_i$.

\section{Decay width for $B^+\to K^+ ~Q\bar{Q}$
  \label{sec:BK}} 
Consider now the $B^+$ meson decay
\be
B^+(\pB)\to K^+ (\pK)\, ~Q(k_1)\, \bar{Q}(k_2)\, ,
\label{BKproc}
\ee
where $\pB$, $\pK$, and $k_{1,2}$ are four-momenta and $Q$ stands for a generic dark-fermion $Q_i$. For simplicity we assume all dark fermions have unit charge $q_i=1$\footnote{This setup serves to approximate more general scenarios where different generations of dark fermions have charges of the same order of magnitude: the overall coupling strength can be reabsorbed in the definition of $\alphaD$.}.

Using the interaction Lagrangian terms in~\eq{Leff} and~\eq{Ldark}, the corresponding decay amplitude is given by
\be
   {\cal M}=\frac{-i e_D}{\ES}
\langle K|[\bar{s} \sigma_{\mu\nu} b]| B\rangle 
\frac{q^{\nu} }{s} \left[\bar{Q} \gamma^{\mu} Q\right]\, ,
\ee
where $s=q^2$ and $q^{\mu}=k_1+k_2$ is the off-shell dark photon momentum.
In order to simplify the notation, in the following we often leave the charge specification for the $B^+$ and $K^+$ mesons understood.

For the tensor hadronic matrix element $\langle K|\bar{s} \sigma_{\mu\nu} b| B\rangle$, we use the parametrization provided in Ref.~\cite{Ali:1999mm}, obtaining 
\be
\langle K(\pK) |\left[\bar{s}\sigma_{\mu\nu} q^{\nu}b\right] |B(\pB)\rangle =
i\Big[(p_B+p_K)_{\mu}s-q_{\mu}(m_B^2-m_K^2)\Big]\frac{f_T(s)}{m_B+m_K}\, ,
\label{BKmatrix}
\ee
where the form factor $f_T(s)$ was computed in Ref.~\cite{Bobeth:2011nj} by using the light cone sum rule (LCSR) approach. Its expression, as a function of $s$, is given by
\be
f_T(s)=\frac{f_T(0)}{1- s/m_{\rm res,T}^2}\left\{
1+b_1^T\left[z(s)-z(0)+\frac{1}{2}\Big(z(s)^2-z(0)^2\Big)\right]\right\}\, ,
\label{K_formF}
\ee
with
\be 
z(s)=\frac{\sqrt{\tau_+-s}-\sqrt{\tau_+-\tau_0}}{\sqrt{\tau_{+}-s}+\sqrt{\tau_{+}-\tau_0}}, ~~\tau_0=\sqrt{\tau_{+}}\left(\sqrt{\tau_{+}}-
\sqrt{\tau_{+}-\tau_{-}}\right), ~~ \tau_{\pm}=\left(m_B\pm m_K\right)^2\, .
\ee
The values of the $b_1^T$ and $f_T(0)$ coefficients, as well as of the resonance mass $m_{\rm res,T}$, can be found in Ref~\cite{Bobeth:2011nj}.
Then, following the notation of Ref.~\cite{Ali:1999mm}, the differential decay width $d\Gamma/d \hat{s}$ for the process $B\to K Q_i\bar{Q}_i$  is found to be
\bea
\frac{d\Gamma^K}{d\hat{s}}&=& \frac{\alphaD m_B^5}{32 \pi^2 m_b^2 \ES^2} \bar{u}(\hat{s})\left(\lambda(\hat{s}) -\frac{\bar{u}(\hat{s})^2}{3}\right)
\frac{\hat{m}_b^2}{(1+\hat{m}_K)^2} |f_T(s)|^2\, ,
\label{dGammaK}
\eea
where $\hat{s}=s/m_B^2$, $\hat{m_b}=m_b/m_B$, $\hat{m_K}=m_K/m_B$, with $m_b$ the $b$ quark pole mass, and $m_B$ and $m_K$ the $B^+$ and $K^+$ masses respectively. The functions $\lambda(\hat{s})$ and $\bar{u}(\hat{s})$ are defined as 
\bea
\bar{u}(\hat{s})&=&\sqrt{\lambda(\hat{s})\Big(1-4 \frac{\hat{m}_Q^2}{\hat{s}}\Big)},\\
\nonumber
\lambda(\hat{s})&=& 1+\hat{m}_K^4+\hat{s}^2-2\hat{s}-2\hat{m}_K^2(1+\hat{s})\, ,
\eea
where $\hat{m}_{Q}=m_{Q}/m_B$, and $m_{Q}$ is the dark fermion mass.

By integrating the differential width in \eq{dGammaK} over the kinematic range of $\hat{s}$,
\be
4 \hat{m}_Q^2 \le\, \hat{s}\, \le (1-\hat{m}_K)^2,
\ee
and dividing by the $B^+$ total width, we obtain the following value for the relevant BR,
\be
   {\rm BR}(B^+\to K^+ Q \bar{Q}) \simeq 5.43 \times 10^{-6}
\left(   \frac{10^5 {\rm TeV}}{\ES}\right)^2 \times \alphaD\, ,
\label{BRK}
\ee
valid for $\hat{m}_Q\simeq 0$.
For comparison, the SM result for the differential width is given by~\cite{Buras:2014fpa}
\be
\frac{d \Gamma^{K}_{\SM}}{d \hat{s}}(B^+\to K^+ \nu \bar{\nu})
= \frac{G_F^2 m_B^5 \alphaEW^2 |V_{tb}V^*_{ts}|^2 X_t^2}{16^2\pi^5\sin^{4}\theta_W}|f_{+}(s)|^2 \lambda^{3/2}(\hat{s})\, ,  
\label{eq:SMgamma}
\ee
where $X_t\simeq 1.469$, $V_{tb}$ and  $V_{ts}$ are the CKM matrix elements,
$\theta_W$ the Weinberg angle and $\alphaEW$ the QED fine structure constant evaluated at the top-quark mass scale. For the input parameters in \eq{eq:SMgamma} we use the central values
\cite{Buras:2014fpa} $\alphaEW^{-1}=127.925$, $\sin^2\theta_W=0.23126$,
$|V_{tb}V^*_{ts}|=0.0401$. The form factor $f_{+}(s)$ is computed in LCSR
approach~\cite{Bobeth:2011nj} and is related to the matrix element of the vector operator as 
\be
    \langle K(\pK) |\left[\bar{s}\gamma_{\mu}b\right] |B(\pB)\rangle
    = f_+(s)\Big[ \big(p_B+p_K\big)_{\mu}-\frac{m_B^2-m_K^2}{s}q_{\mu}\Big] \nonumber \\
    +\frac{m_B^2-m_K^2}{s}f_0(s)q_{\mu}\, .
    \ee
Its expression is given by~\cite{Bobeth:2011nj}  
\be
f_+(s)=\frac{f_+(0)}{1- s/m_{\rm res,+}^2}\left\{
1+b_1^+\left[z(s)-z(0)+\frac{1}{2}\Big(z(s)^2-z(0)^2\Big)\right]\right\}\, ,
\label{BK_formSM}
\ee
where the values for the coefficients $b_1^+$ and resonant mass $m_{\rm res,+}$
can be found in Ref.~\cite{Bobeth:2011nj}.  Notice that, for vanishing dark fermion masses, the $q^2$ distributions in Eq.~\eqref{dGammaK} tend to the SM contribution modulo an overall multiplying factor.

\subsection{Sommerfeld-Fermi and dark magnetic-dipole corrections}

The Sommerfeld-Fermi enhancement (SF)~\cite{Sommerfeld:1931qaf,Fermi:1934hr} is a long-distance correction to the amplitude in~\eq{dGammaK} relevant for the large values of $\alphaD$ required to explain the excess. The re-summation of the leading $\log$ terms induced, in this case, by the soft photon-like corrections, is equivalent to the inclusion of the Coulomb interaction in the wave function of final states~\cite{Weinberg:1965nx}. In our case, the contribution is particularly important in a kinematic regime where the final charged particles are non-relativistic in the rest frame of the decaying $B$ meson. 

The SF correction can be accounted for in the computation of the differential branching ratio by including the corresponding overall enhancement factor~\cite{Isidori:2007zt}
\bea
d\Gamma^K_{\SF}&=&
\Omega(\beta_{12})\, d\Gamma^K\,,\label{SFen}\\ \nonumber
\Omega(\beta_{12})&=&\frac{2\pi\alphaD q_i^2}{\beta_{12}}
\frac{1}{1-\exp{\left[-\frac{2\pi\alphaD q_i^2}{\beta_{12}}\right]}}\,, 
\eea
where $d\Gamma^K$ is given in \eq{dGammaK} and
\be
\beta_{12}=\sqrt{1-\frac{m_Q^4}{(s-2 m_Q^2)^2}},
\ee
with $q_i$ being the $U(1)_D$ charge of the dark fermion of the  generation $i$ (that we set to unity in our analysis). For scenarios with large values of $\alphaD \simeq {\cal O}(1)$, the SF provides large contributions also in the $q^2$ region away from the dark fermion production threshold. Furthermore, in the presence of this correction, the dependence of the differential BR on $\alphaD$ is more involved than in~\eq{dGammaK}, where the parameters simply gives an overall multiplying factor.

\begin{figure}[t]
  \centering
  \includegraphics[width=0.85\linewidth]{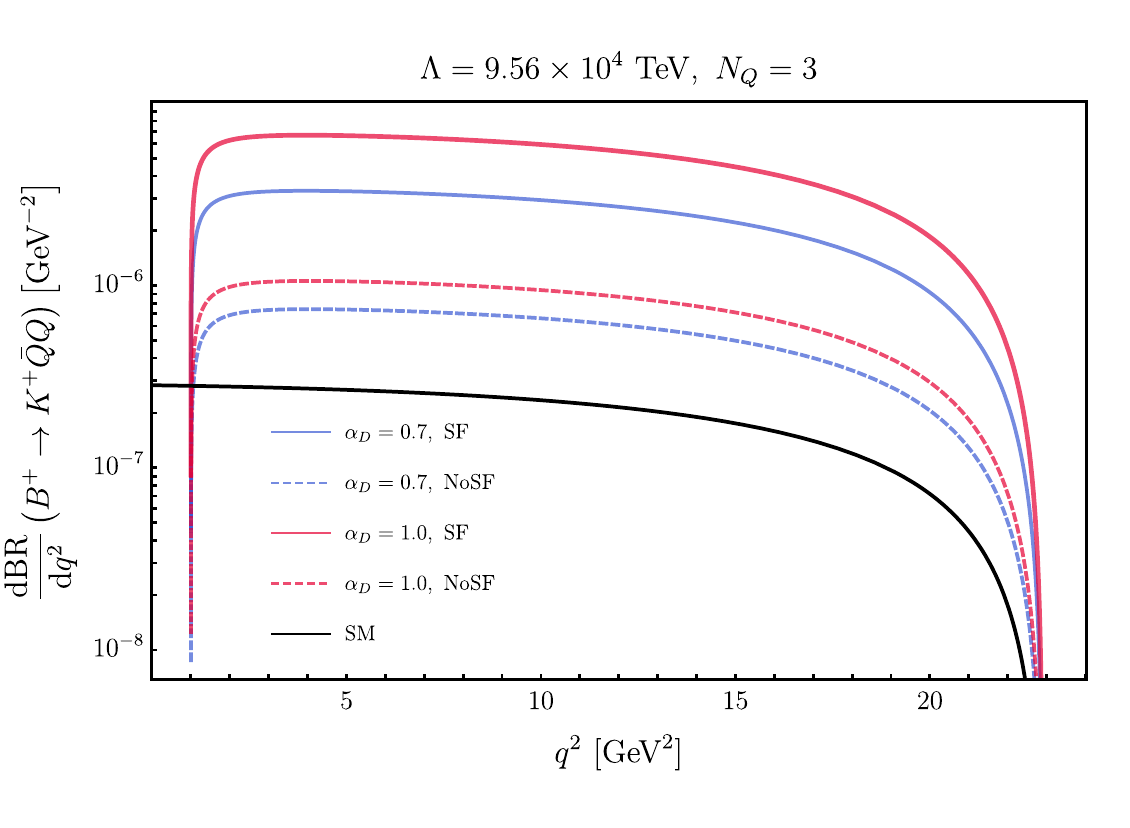}
  \caption{Differential BR for the process $B^+\to K^+ ~Q\bar{Q}$ as a function of the missing energy $q^2$, for $N_Q=3$ generations of dark fermions all with mass $m_Q=0.5$ GeV. The dashed lines use the expression in~\eq{dGammaK}, whereas the solid lines include the SF correction, \eq{SFen}. The SM result, in~\eq{eq:SMgamma}, is shown in black.}
  \label{fig:brs}
\end{figure}

The effect of SF corrections is shown in Fig.~\ref{fig:brs}, where we plot the dependence of the relevant differential BR on $q^2$ for three generations of dark fermions and different values of the NP parameters, including (solid lines) and omitting (dashed lines) the SF enhancement. The SM result, included for reference, is shown in black. As we can see, the effect of the SF correction is quite large: over the relevant kinematic region, the SF correction results in an enhancement of the differential BR by a factor of about 5 for the examined cases of $\alphaD = 0.7$ and $\alphaD = 1.0$.

Besides the SF enhancement we scrutinize also the impact of radiative corrections sourced by the dark $U(1)$ interaction. As the SM particles do not partake in it, the corrections affect only the diagram involving dark fermions. These include, at the one-loop level, the dark photon - dark fermion vertex and the self-energy diagram of the latter. Since we are interested in long-distance effects, we retain only the  ${\cal O}(\alphaD)$ dark magnetic-dipole corrections affecting the $F_2$ form factor -- which multiplies the $\sigma_{\mu\nu}$ operator in the dark fermion effective vertex. The correction is relevant for large values of $\alphaD$ in the $q^2$ region close to the kinematic threshold and can be included in the computation by replacing the function $\lambda(\hat{s})$ in Eq.~\eqref{dGammaK} with 
\be
\lambda(\hat{s})\to \lambda(\hat{s}) \Big[1+\Delta_M(\hat{s})\Big]\, ,
\ee
where the correction $\Delta_M(\hat{s})$ is given by~\cite{Gabrielli:2019uqu}
\be
\Delta_M(\hat{s})=4\hat{m}_Q \,{\rm Re}
\left[F_2(\hat{s})\right]
+\frac{1}{3}\left|F_2(\hat{s})\right|^2\left(\hat{s}+8\hat{m}_Q^2\right)\, ,
\ee
and the magnetic form factor $F_2(\hat{s})$  is
\bea
F_2(\hat{s})&=&\frac{\alphaD}{2\pi}\frac{\hat{m}_Q}
  {\sqrt{\hat{s}(\hat{s}-4\hat{m}_Q^2)}}\log{\left[\frac{2\hat{m}_Q^2-\hat{s}+\sqrt{\hat{s}(\hat{s}-4\hat{m}_Q^2)}}{2\hat{m}_Q^2}\right]}\, ,
\eea
with $\log$ the logarithm in the natural basis. Differently from the case of the SF correction, we find that the integrated effect of the magnetic-dipole correction on the $q^2$ region relevant for the Belle II excess, being at the percent level, is negligible.

\section{Decay width for $B^0\to K^* ~ \gammaD$
  \label{sec:BKstar}} 
As mentioned in the introduction, the presence of a dark photon can also affect other decay processes involving the $b\to s$ FCNC transition. In particular, angular momentum conservation allows the emission of an on-shell massless dark photon in the following decay of the $B^0$ neutral meson
\be
B^0(p_B)\to K^*(p_K) ~ \gammaD(q)\, ,
\label{BKsproc}
\ee
where $p_B$, $p_K$, and $q$ indicate the involved four momenta. The process, which proceeds analogously to the $B^0\to K^* ~ \gamma$ decay in the SM, is induced by the interaction Lagrangian in \eq{Leff} used to model the decay in~\eq{BKproc}.

The matrix element for this transition is given by
\be
   {\cal M}=\frac{-i e_D}{\ES}
   \langle K^*|\left[\bar{s} \sigma_{\mu\nu}q^{\nu} b\right]| B^0\rangle  \epsilonD^{\mu}(q)\, ,
\ee
where $\epsilonD^{\mu}(q)$ is the dark photon polarization vector. For the tensor hadronic matrix element $\langle K^*|\bar{s} \sigma_{\mu\nu} b| B^0\rangle$ we use again the parametrization provided in Ref.~\cite{Ali:1999mm}, which in the general case of $q^2\neq 0$ gives
\bea
\langle K^*(\pK) |\left[\bar{s}\sigma_{\mu\nu}q^{\nu}b\right] |B^0(\pB)\rangle &=& i2\epsilon_{\mu\nu\alpha\beta}\, \varepsilon^{\dag \nu} \pB^{\alpha}\pK^{\beta} T_1(s)\nonumber\\
&+&T_2(s)\left[\varepsilon^{\dag}_{\mu}\left(\mB^2-\mKs^2\right)-
  \left(\varepsilon^{\dag}\cdot \pB\right)\left(\pB+\pK\right)_{\mu}\right]
\nonumber\\
&+&T_3(s)\left(\varepsilon^{\dag}\cdot \pB\right)\left[q_{\mu}-
  \frac{s}{\mB^2-\mKs^2}\left(\pB+\pK\right)_{\mu}\right]\, ,
\label{KstarME}
\eea
where $\varepsilon$ and $m_{K^*}$ are the polarization vector and mass of the $K^*$ meson, respectively, and $m_B$ is the $B^0$ mass. The terms $T_{1,2,3}(s)$ are form factors and the convention $\epsilon_{0123}=+1$ is used for the Levi-Civita symbol. Only the $T_1(s)$ form factor contributes to the matrix element of \eq{KstarME} for the operator $[\bar{s} \sigma_{\mu\nu}q^{\nu} b]$, while
$T_{2,3}(s)=0$  due to parity.

By combining the effective Lagrangian in Eq.~\eqref{Leff} with the hadronic matrix elements in Eq.~\eqref{KstarME}, we obtain for the corresponding decay width~\cite{Gabrielli:2019uqu}
\bea
\Gamma(B^0\to K^* \gammaD) &=& \frac{m_B^3\, |T_1(0)|^2\left(1-\hat{m}^2_{K^*}\right)^3}  {32\pi \ES^2}\, ,
\eea
where $\hat{m}_{K^*}=m_{K^*}/m_B$ and $T_{1}(0)\simeq 0.28$ \cite{Bobeth:2011nj} is the hadronic form factor evaluated at $s=0$ and defined in Eq.~\eqref{KstarME}.

Finally, the BR of the process under scrutiny then is
\be
{\rm BR}(B^0\to K^* \gammaD)= 2.47 \times 10^{-5} \left(\frac{ 10^5 {\rm TeV}}{\ES}\right)^2\, .
\label{BRKs}
\ee

Present searches from the BaBar~\cite{BaBar:2013npw} and Belle~\cite{Belle:2017oht} collaborations have only set upper bounds on the BR$(B^0\to K^* ~\nu \bar{\nu})$ which is compatible with the SM expectations~\cite{Altmannshofer:2009ma}. In particular, the most recent result by the Belle collaboration is ${\rm BR}(B^0\to K^* \nu \bar{\nu}) < 2.7\times 10^{-5}$ at 90\% C.L.~\cite{Belle:2017oht}. The invisible $\nu\bar{\nu}$ system shows as continuous missing energy $E_{\rm miss} $ distribution in the detector and a kinematic cut $E_{\rm miss} > 2.5 {\rm GeV}$, in the $B^0$ meson rest frame, has been used in the analysis.

It could be thought that the upper bound on ${\rm BR}(B^0\to K^* \nu \bar{\nu})$ may be used to constrain also the $B^0\to K^* \gammaD$ decay. However, the experimental signatures of the two processes differ in that the final state possesses different kinematic properties. In more detail, the two-body decay $B^0\to K^* \gammaD$ yields a monochromatic missing energy distribution set at the specific value $E_{\gammaD}=E_{\rm miss}=2.56\, {\rm GeV}$ in the $B^0$ rest frame. As mentioned before, three bodies decays such as $B^0\to K^* \nu \bar{\nu}$ result instead in a continuous $E_{\rm miss}$ distribution. For this reason, strictly speaking, applying the limits in Ref.~\cite{Belle:2017oht} to the ${\rm BR}(B^0\to K^* ~\gammaD)$ is not completely correct and a dedicated analysis would be required to set limits on the process~\cite{Belle-pcomm}. Bearing this caveat in mind, in the following we will still assume that the Belle upper bound on the ${\rm BR}(B^0\to K^* \nu \bar{\nu})$ holds true also for the $B^0\to K^* \gammaD$ decay and analyze its phenomenological implications within the present context.

Imposing the experimental upper limit on  ${\rm BR}(B^0\to K^* \nu \bar{\nu}) < 2.7\times 10^{-5}$ gives a lower bound,
\bea
\ES \geq 9.56 \times 10^4\, {\rm TeV},
\label{Bellebounds}
\eea
that is about 32 times stronger than the one obtained from the inclusive decay
${\rm BR}(b\to s X_{\rm inv})<{\cal O}(10\%)$~\cite{Ciuchini:1996vw}. 

We remark that the bound in \eq{Bellebounds} allows the framework to satisfy also the NP constraint due to the $B\to X_s \gamma$ decay at 95\% C.L. even when large values of $\alphaD$ are considered~\cite{Gabrielli:2019uqu}. The relevance of this latter bound, however, strongly depends on the UV completion  of the theory describing the emergence of the effective scale $\ES$.

\section{Fitting the $B^+\to K^+ \nu \bar{\nu}$ excess}
\label{sec:BKanom}

The Belle II collaboration has recently provided a measurement of the  $B^+\to K^+ \nu \bar{\nu}$ decay observed with a significance of 3.5$\sigma$. Combining the two methods adopted in this search gives~\cite{Belle-II:talk1,Belle-II:2023esi}
\be
{\rm BR}(B^+\to K^+ \nu \bar{\nu})=(2.3 \pm 0.7)\times 10^{-5}\, ,
\ee
to be compared with the SM prediction 
\be
{\rm BR}(B^+\to K^+ \nu \bar{\nu})=(4.29 \pm 0.23)\times 10^{-6}\, .
\ee
We assume that the difference is due to NP effects
\be
{\rm BR}(B^+\to K^+ + {\rm inv})_{\rm NP}=(1.9 \pm 0.7)\times 10^{-5}\,,
\label{eq:excess}
\ee
with the missing energy being carried away by the dark fermion in place of SM neutrinos.

In order to simplify the analysis, we fit the excess in~\eq{eq:excess} assuming the production of $N_Q$ generations of dark-fermions all with unit charge and consider the inclusive decay defined as
${\rm BR}(B^+\to K^+  Q_X \bar{Q}_X)= \sum_{i=1}^{N_Q} \left[{\rm BR}(B^+\to K^+  Q_i\bar{Q}_i)\right]$. Accounting for the constraint discussed in the previous section and for the SF correction we have
\be
\frac{{\rm BR}(B^0\to K^* \gammaD)}{{\rm BR}(B^+\to K^+ Q_X \bar{Q}_X)_{\SF}}\simeq \frac{0.72}{N_Q}\qquad{\text{for $\alphaD\sim {\cal O}(1)$},}
\ee
and imposing ${\rm BR}(B^+\to K^+ Q_X \bar{Q}_X)={\rm BR}(B^+\to K^+ + {\rm inv})_{\rm NP} \sim 1.9 \times 10^{-5} $ then gives
\be
{\rm BR}(B^0\to K^* \gammaD)\simeq \frac{1.35}{ N_Q}\times 10^{-5}\,.
\ee
Since no excess has been observed for the $B^0\to K^* \nu \bar{\nu}$ transition, applying the upper bound on ${\rm BR}(B^0\to K^* \nu \bar{\nu}) < 2.7\times 10^{-5}$ to the ${\rm BR}(B^0\to K^* \gammaD)$ we obtain 
\be
   N_Q \ge  1\, ,
\label{alphaDbound}
\ee
for the number of dark fermion generations. \eq{alphaDbound} then shows that a strong (but still perturbative) dark photon coupling to the dark sector, $\alphaD\sim {\cal O}(1)$, allows to explain the excess in the decay
${\rm BR}(B^+\to K^+ \nu \bar{\nu})$ through the incoherent production of $N_Q$ light dark fermions. 

\begin{figure}
  \centering
  \includegraphics[width=.75\linewidth]{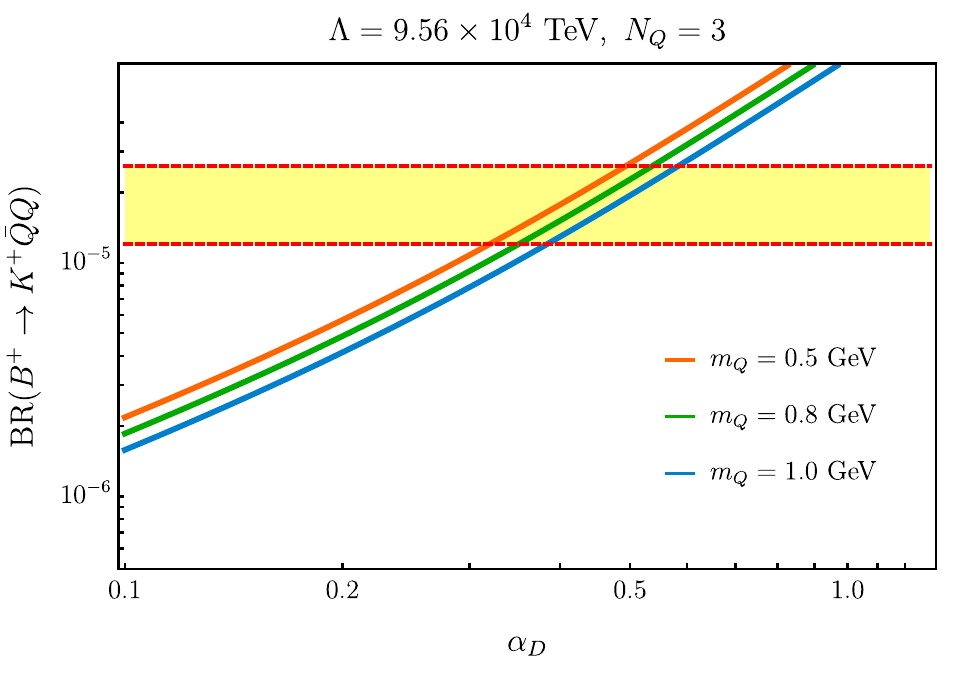}
  \\
  \includegraphics[width=.75\linewidth]{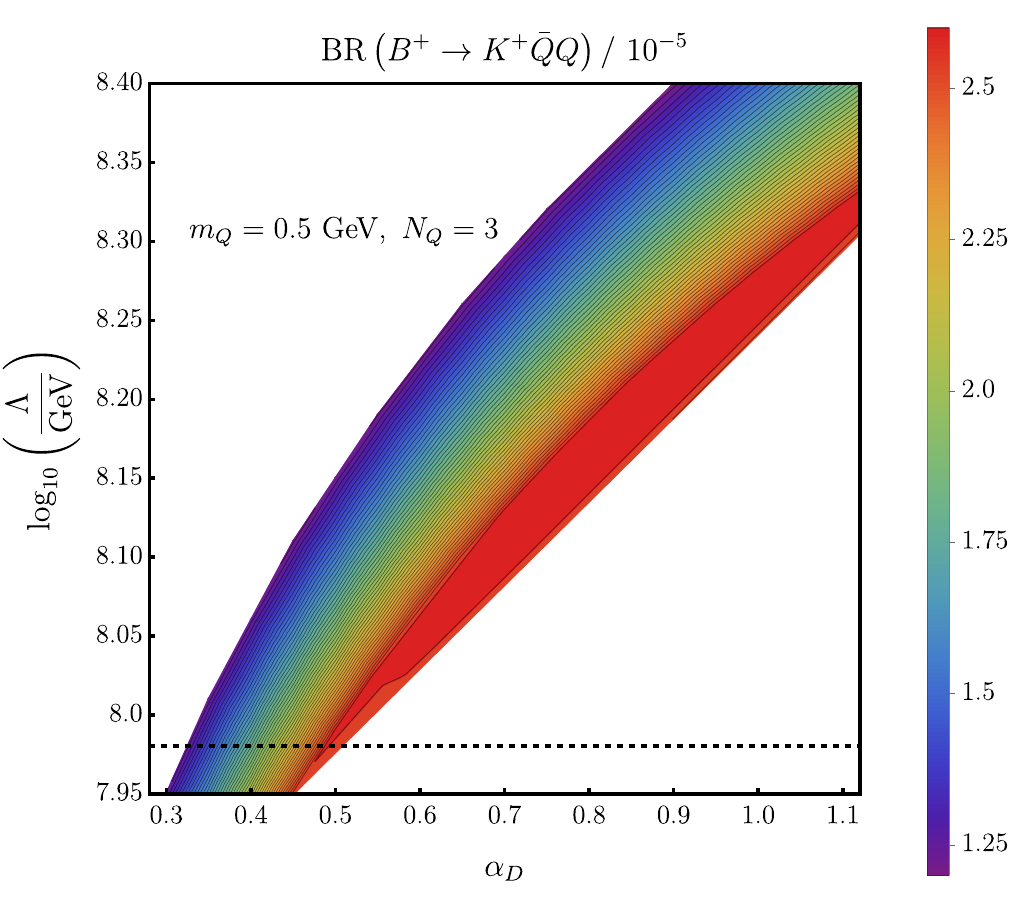} 
  \caption{Values of the parameters for which the scenario can explain the Belle II excess through the BR given by~\eq{SFen}. The top panel shows three different solutions obtained by varying the mass value shared by three generation of dark fermions for the value of the effective scale $\ES$ indicated by~\eq{Bellebounds}. The bottom panel shows the dependence on the effective scale and dark fine-structure constant $\alphaD$ of the BR obtained by integrating~\eq{SFen}.   }
  \label{fig:pars}
\end{figure}

In Fig.~\ref{fig:pars} we show the parameter values for which our scenario can reproduce the Belle II excess through~\eq{SFen}. The yellow horizontal band in the top panel shows the $1\sigma$ confidence interval identified by the experiment. The colored lines show the solutions of the model for the indicated values of the dark fermion masses, assuming three generations of particles degenerate in mass. As we can see, the anomaly can be matched for perturbative values of the (universal) dark fine-structure constant $\alphaD$. In the plot we used the value of the effective scale $\ES$ saturating the lower bound in~\eq{Bellebounds}. The bottom panel shows instead the value of the BR relevant to reproduce the excess as a function of the same scale and $\alphaD$ for a benchmark value of the dark fermion mass, $m_Q=0.5$ GeV, shared by three generations of particles. The constraint in~\eq{Bellebounds} excludes the area below the black dashed line.      

The relation that the scenario establishes between the decays $B^0\to K^* \gammaD$ and $B^+\to K^+ Q_X \bar{Q}_X$ is explored in Fig.~\ref{fig:fig3}. The solid lines show the ratio of the corresponding BRs as a function of $\alphaD$, obtained for $N_Q=3$ generation of dark fermions sharing the indicated mass values. Importantly, the dependence of the BRs on the effective scale $\ES$ drops out in the ratio. The red dashed line shows the value supported by the NP interpretation of the Belle II excess, \eq{eq:excess}, once the upper bound ${\rm BR}(B^0\to K^* \nu \bar{\nu}) < 2.7\times 10^{-5}$ is assumed. The latter excludes the solutions falling in the area dashed in red. Future measurements of ${\rm BR}(B^0\to K^* \nu \bar{\nu})$ can be used to constrain the value of the dark fine-structure constant $\alphaD$ and a simultaneous fit of the Belle II excess would then allow the determination of the effective scale $\ES$. 

\begin{figure}[h]
  \centering
  \includegraphics[width=.75\linewidth]{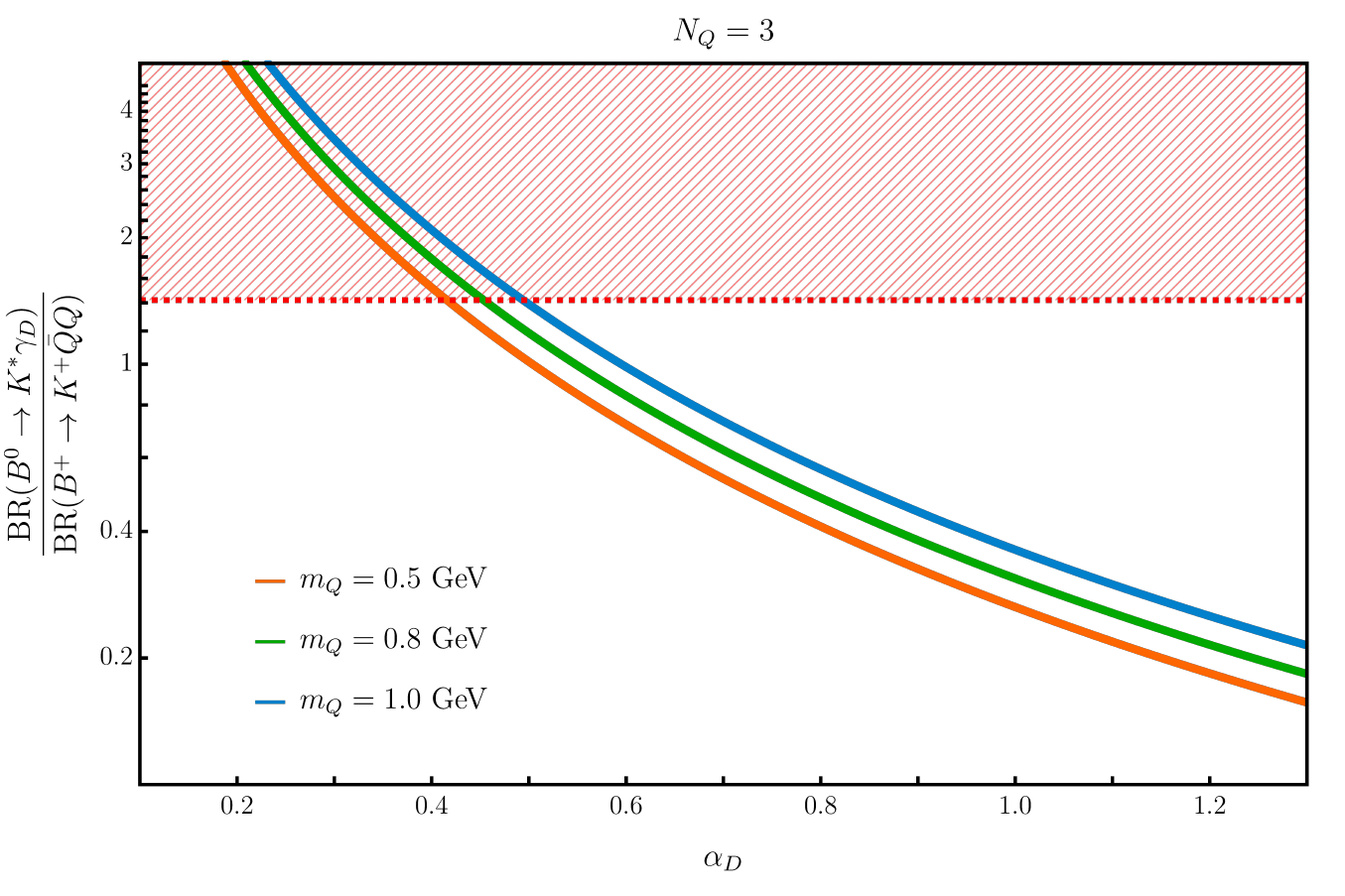}
  \caption{Dependence of ${\rm BR}(B^0\to K^* \gammaD)$/${\rm BR}(B^+\to K^+ Q_X \bar{Q}_X)$ on the dark fine-structure constant $\alphaD$. The ratio is for $N_Q=3$ generations of dark fermions sharing the indicated mass values and does not depend on the effective scale $\ES$. The dashed red line indicates the value obtained for the NP interpretation of the Belle II excess, \eq{eq:excess}, assuming the current experimental bound ${\rm BR}(B^0\to K^* \nu \bar{\nu}) < 2.7\times 10^{-5}$. The latter excludes solution falling in the area hatched in red. Future determinations of ${\rm BR}(B^0\to K^* \nu \bar{\nu})$ can be used to constrain the parameter space of the model as described in the text. }
  \label{fig:fig3}
\end{figure}

\section{Predicting a possible signal in $B^0_s$ decays
  \label{sec:fpredict}}
In the previous section we have shown that the Belle II excess in
$B^+\to K^+\nu \bar{\nu}$ can be explained by the production of dark fermion pairs in the process $B^+\to K^+ Q_i \bar{Q}_i$, mediated by a massless off-shell dark photon. In order to test this interpretation of the Belle II data, we seek further phenomenological consequences of the same interaction Lagrangian~\eqref{Leff} used to model the excess.

The magnetic-dipole operator in \eq{Leff} for the $b\to s$ transition can also contribute to the $B_s^0\to \phi  \nu \bar{\nu}$ decay, thereby inducing an excess over the SM predictions for $B_s^0\to \phi \nu \bar{\nu}$~\cite{Colangelo:2007jy}
\be
{\rm BR}(B^0_s\to \phi  \nu \bar{\nu})_{\rm SM}=(1.4\pm 0.4)\times 10^{-5}\, .
\ee
The process $B_s ^0\to \phi  \, \gammaD $ shows in experiments as $B_s^0\to \phi + {\rm E_{miss}}$ with the massless dark photon carrying away an energy $E_{\rm miss}=2.59 {\rm GeV}$ in the $B_s^0$ rest frame, in complete analogy with the $B^0\to K^* \gammaD$ transition. The present upper limit on this BR comes from LEP data and is given by
\cite{DELPHI:1996ohp}
\be
{\rm BR}(B_s^0\to \phi  \nu \bar{\nu})_{\rm exp} <5.4 \times 10^{-3}
\ee
at 90\% confidence level.

The prediction of the scenario for the ${\rm BR}(B_s^0 \to \phi  \gammaD)$ can be obtained by noticing that the transition depends on the same hadronic matrix element (the magnetic dipole operator $\langle \phi|[s\sigma_{\mu\nu}b]|B^0_s\rangle$) that enters the $B^0_s \to \phi  \gamma$ decay. The same holds true for the matrix element $\langle K^*|[s\sigma_{\mu\nu}b]|B^0\rangle$, which enters both the  ${\rm BR}(B^0 \to K^*  \gammaD)$ and ${\rm BR}(B^0 \to K^* \gamma)$. The ratio measured at LHCb~\cite{LHCb:2012quo}
\be
R_{K^* \phi}=\frac{{\rm BR}(B^0\to K^* \gamma)}{{\rm BR}(B_s^0\to \phi \gamma)}
=1.23 ~(6)_{\rm stat}~(4)_{\rm syst} ~ (10)_{f_s/f_d}\, ,
\ee
where the third uncertainty is associated with the ratio of fragmentation
fractions $f_s /f_d$,  allows us to infer the hadronic tensor form factor
$T_1^{B_s^0 \to \phi}$ for the ${\rm BR}(B_s^0 \to \phi  \gammaD)$---or, analogously, for  ${\rm BR}(B_s^0 \to \phi  \gamma)$---
from the corresponding tensor form factor $T_1^{B^0 \to K^* }$
of ${\rm BR}(B^0 \to K^* \gammaD)$ in \eq{KstarME} evaluated at $s=0$. In particular, Ref.~\cite{Lyon:2013gba} gives
\be
R_{K^*\phi}=|r_{K^*\phi}|^2 C_{K^* \phi}\left(1+\delta_{WA}\right)\, ,
\label{RKsphi}
\ee
where the term $\delta_{WA}\simeq 0.02$ contains weak interaction corrections, 
\be
r_{K^*\phi}=\frac{T_1^{B^0\to K^*}}{T_1^{B^0_s \to \phi}}\, ,
\ee
and 
\be
C_{K^* \phi}=\frac{\tau_{B^0}}{\tau_{B^0_s}}\left(\frac{m_B}{m_{B_s}}\right)^3
\left(\frac{1-m_{K^*}^2/m_{B}^2}{1-m_{\phi}^2/m_{B_s}^2}\right)^3=1.01\, .
\ee
Since $\delta_{WA}$ is negligible, we approximate $R_{K^*\phi}\simeq r_{K^*\phi}^2$ and with the central value from \eq{RKsphi} extract $r_{K^*\phi}$ to predict the ${\rm BR}(B_s^0 \to \phi  \gammaD)$ to be
\be
{\rm BR}(B_s^0\to \phi \gammaD)=4.02 \times 10^{-5} \left(\frac{ 10^5 {\rm TeV}}{\ES}\right)^2\, .
\label{BRphi}
\ee
Finally, by using the value of the scale $\ES$ saturating the lower bound in~\eq{Bellebounds} we obtain 
\be
  {\rm BR}(B^0_s\to \phi  \gammaD)\lsim 2\times 10^{-5}\, ,
\label{BRphi2}
\ee
where, importantly, the upper bound falls within the reach of dedicated LHCb searches.

\section{Conclusions
  \label{sec:concl}}
We have shown that the recent measurement of  BR$(B^+\to K^+ \nu \bar{\nu})$ by the Belle II Collaboration, yielding a $2.7\sigma$ discrepancy with respect to the SM prediction, can be explained by dark fermion pair production through the process $B^+\to K^+ Q_i \bar{Q}_i$ mediated by an off-shell massless dark photon. The excess can be fitted with dark fermion masses of the order of 500 MeV, which experimentally result in a  missing energy signature. The absence of the two body decay $B^+\to K^+\gammaD$ is ensured, for a massless dark photon, by the angular momentum conservation.   

Since a dark photon emission is allowed in the $B^0\to K^* \gammaD$ transition, the lack of an observed signal related to the  $B^0\to K^* \nu \bar{\nu}$ decay gives $N_Q\gsim 1$ for $\alphaD\sim {\cal O}(1)$. In regard of this, large but perturbative values of $\alphaD$ are still experimentally viable for a massless dark photon, as the scenario is only loosely constrained due to the naturally suppressed couplings with the visible sector. We also remark that the bound for the $B^0\to K^* \nu \bar{\nu}$ process cannot be straightforwardly taken at face value to constrain the scenario via the related $B^0\to K^*  \gammaD$ due to differences in the kinematics of the involved final states. 

If the proposed dark fermion production mechanism is responsible for the Belle II excess, we predict the appearance of NP signals in the $B_s^0\to \phi \gammaD$ decay (and in $B^0\to K^* \gammaD$, should the aforementioned bound not hold), which could be scrutinized at the next LHCb experiment. These searches are crucial to test the massless dark photon interpretation of the Belle II excess in $B^+\to K^+ \nu \bar{\nu}$ decays. The expected phenomenological signature, in both channels, is a monochromatic missing energy distribution centered at the value $E_{\rm miss}=2.56\, {\rm GeV}$ and $E_{\rm miss}=2.59\, {\rm GeV}$ in the B meson rest frame, for the $B^0\to K^* \gammaD$ and  $B_s^0\to \phi \gammaD$ transitions, respectively. We urge the experimental collaborations to start dedicated analysis targeted at these decay modes. 

\section*{Acknowledgements}
We thank Alexander Glazov and Diego Tonelli for useful discussions.
This work was supported by the Estonian Research Council grants PRG803, RVTT3, RVTT7 and by the CoE program grant TK202 ``Fundamental Universe'’.

\bibliographystyle{JHEP}

\bibliography{BKnunu_hep.bib}

\end{document}